\documentclass[11pt,twoside]{article}
\usepackage{amssymb}
\usepackage{latexsym}
\usepackage[dvips]{graphicx}
\usepackage{color}

\newcommand{\boldgreek}[1]{\mbox{\boldmath$#1$}}

\newcommand{\R}{I\kern-0.37emR}

\newcommand{\ny}{n\rightarrow\infty}
\newcommand{\Q}{I\kern-0.37emP}
\newcommand{\E}{I\kern-0.37emE}

\date{}
\title{{Estimation of Quantile Functionals in Linear Model}\\[2mm]
\textit{\small Devoted to the Memory of Pranab Kumar Sen}}
\author{\textsc{Jana Jure\v{c}kov\'a, Jan Picek\thanks{The research of Jureckova and Picek was supported by 
the Grant GA\v{C}R 22-03636S}, Jan Kalina}\\[2mm]
{\it \normalsize{The Czech Academy of Sciences, Institute of Information Theory and Automation}},\\ 
{\it \normalsize {Technical University in Liberec}},\\ {\it \normalsize {The Czech Academy of Sciences, Institute of Computer Science, Prague}}}
\begin{document}
\maketitle
\thispagestyle{empty}


\begin{abstract}
Various indicators and measures of the real life procedures rise up as  functionals of the quantile process of a parent random variable $Z$.  However, $Z$  
can be observed only through a response in a linear model whose covariates are not under our control and the probability distribution of error terms is generally 
unknown. The problem is that of nonparametric estimation or other inference for such functionals. We propose an estimation procedure based on the 
\textit{averaged two-step regression quantile}, recently developed by the authors, combined with an R-estimator of slopes of the linear model. 
\end{abstract}

\section{Introduction}
\setcounter{equation}{0}
Various indicators and measures of the real life procedures rise up as  functionals of the quantile process of a parent random variable $Z$. 
We assume that $Z$ has an absolutely continuous distribution function $F$ and quantile function $Q(\alpha)=F^{-1}(\alpha)  ; 0<\alpha<1,$
but their forms are generally unknown. The functionals of $Q$ can be 
environmental, industrial, health, physical, economic  and other indicators. To study and predict their behavior,  we should monitor 
the values of $Z$ over time or space. However, the values of  $Z$
often cannot be directly observed; they can be  hidden and not directly physically measurable,  or can be autocorrelated 
or regressed with covariates which are not under our control. 

We shall illustate this situation on estimation of  a functional of the quantile function $Q$ of  parent random variable $Z,$ which is observable 
only through a response $Y$
of the linear regression model, with regressors of unknown intensities. Hence, instead of observations $\mathbf Z_n = 
(Z_{1},\ldots,Z_{n})^{\top}$, we can observe only the
variables $\mathbf Y_n = (Y_{n1}, . . . , Y_{nn})^{\top}$, what are the $\mathbf Z_n$
affected by covariates $\mathbf X_n$ with unknown intensities, measured by regression 
coefficients $\beta_1, \ldots,\beta_n.$ 
Summarizing, we work with the linear regression model
\begin{equation}
\label{1}
{\bf Y}_n=\beta_0\mathbf 1_n+{\mathbf X}_n{\boldgreek\beta}+{\mathbf Z}_n
\end{equation}
with observations ${\mathbf Y}_n=(Y_{n1},\ldots,Y_{nn})^{\top},$ unknown parameters $\beta_0$ (intercept),\\ 
${\boldgreek\beta}=(\beta_1,\beta_2,\ldots,\beta_p)^{\top}$ (scales), $\mathbf 1_n=(1,\ldots,1)^{\top}\in\R_n$
and the $n\times p$ matrix ${\mathbf X}={\mathbf X}_n$ of covariates. It is a known matrix with the rows 
$\mathbf x_i^{\top}=(x_{i1},\ldots, x_{ip}),  \; i=1,\ldots,n.$
Because $Z_1,\ldots, Z_n$ are not available, every inference on the functionals of $Q$ 
 is possible only by means of observations of $Y$. 
The variables $Z_1,\ldots, Z_n$ are  assumed to be independent and identically distributed (\textit{i.i.d.})  and
\begin{equation}\label{df}
\int zdF(z)=0,\quad \int z^2dF(z)<\infty,
\end{equation}
but the methods  can be extended to  autocovariated and sequentially dependent observations. 

We are interested in estimating  the value of the functional $\mathcal S(Q)$ of the quantile function $Q.$ That can be a risk measure, energy consumption 
over a period, water flow over a period, etc. Our interest is also in the confidence interval for $\mathcal S(Q),$ or in a comparison of two functionals, 
representing two possible 
treatments. Our main tool for model (\ref{1}) are the two-step averaged regression quantiles, combined with the rank-estimators (R-estimators) of the slope 
components $\boldgreek\beta.$  These concepts and methods have been elaborated  by the authors (\cite{Averaged}, \cite{Sankhya2005}, \cite{Aplimat}, 
\cite{Sen}) , following  the pioneering ideas of Koenker and Bassett (1978).    \cite{Pflug}, 
\cite{Bassett} and \cite{Trindade} use the quantile regression for estimating the conditional value at risk. 

It is shown in \cite{Aplimat} that the averaged two-step regression $\alpha$-quantile, introduced in \cite{Averaged}, approximates the quantile 
function $Q(\alpha)$ of $Z$, asymptotically in probability for $\ny$, up-to the regression parameters.  

In the present paper, we deal with  estimating the linear functional of the quantile function  of the main variable in the linear regression model, in the situation 
that this variable  is unobservable, affected by the covariates. 
Our estimator 
is based 
on  the averaged  two-step regression $\alpha$-quantile, combined with R-estimator of the scale parameters of the model. 
As it is shown by authors in \cite{Aplimat}, the averaged two-step regression $\alpha$-quantile, introduced in \cite{Averaged}, approximates the quantile 
function $Q(\alpha)$ of the errors, asymptotically in probability for $\ny$, up-to the regression parameters. 
Because its number of breakpoints equals exactly to $n$, while in the case of ordinary regression quantile it is much larger, and it is nondecreasing in 
$\alpha\in (0,1),$ the averaged two-step regression quantile 
also  facilitates  joint fitting several quantile functionals in the regression model.

\section{Regression quantile and its two-step version}
\setcounter{equation}{0}
\subsection{Regression quantile}
The model (\ref{1}) can be  rewritten as the model
\begin{equation}\label{1b}
Y_{ni}=\beta_0+{\mathbf x}_{ni}^{\top}{\boldgreek\beta}+Z_{ni}, \;  i=1,\ldots,n
\end{equation} 
with covariates $\mathbf x_{n1},\ldots, \mathbf x_{nn},$ each element of $\R_p.$ For the sake of brevity, we also use the notation
 $\mathbf x^*_{ni}=(1,x_{i1},\ldots,x_{ip})^{\top}, \; i=1,\ldots,n.$
Let $\widehat{\boldgreek\beta}_n(\alpha)\in R_{p+1}$ be the $\alpha$-regression quantile of model (\ref{1}), $0<\alpha<1$, 
i.e. the solution of the minimization
\begin{equation}\label{RQ}
\sum_{i=1}^n\rho_{\alpha}(Y_i-b_0-\mathbf x_i^{\top}\mathbf b)=\min, \; b_0\in\R_1, \; \mathbf b\in\R_{p}.
\end{equation}
If  derivative $f$ of $F$ exists and is positive
in a neighborhood of the quantile $Q(\alpha)=F^{-1}(\alpha),$
and if the matrix $${\mathbf D}_n=n^{-1}\left(\mathbf 1_n, \mathbf X_n\right)^{\top}\left(\mathbf 1_n, \mathbf X_n\right)$$
is positively definite starting with some $n$, then
$n^{\frac 12}(\widehat{\boldgreek\beta}_n(\alpha)-%
\check{\boldgreek\beta}(\alpha))$ admits the asymptotic representation (see e.g. \cite{Sen})
\begin{equation}
\label{3a}
n^{\frac 12}(\widehat{\boldgreek\beta}_n(\alpha)-\check{\boldgreek\beta}(\alpha))=
n^{-\frac 12} (f(F^{-1}(\alpha))^{-1}{\mathbf Q}_n^{-1}\sum_{i=1}^n{\mathbf x}_i^*(\alpha-I[Z_i<F^{-1}(\alpha)])+o_p(1)
\end{equation}
as $n\rightarrow\infty,$ where $\check{\boldgreek\beta}(\alpha)=(F^{-1}(\alpha)+\beta_0,\beta_1,\ldots,\beta_p)^{\top}$ is the population counterpart
 of the regression quantile. 
The intercept part of the representation (\ref{3a}) is rewritten as
\begin{eqnarray}
\label{4}
\hat{\beta}_{n0}(\alpha)-\beta_0-Q(\alpha)&=&(nf(Q(\alpha)))^{-1}\sum_{i=1}^n(\alpha-I[Z_i<F^{-1}(\alpha)])+o_p(n^{-\frac 12})\nonumber\\
&=&Z_{n:[n\alpha]}-Q(\alpha)+o_p(n^{-\frac 12})
\end{eqnarray}
as $n\rightarrow\infty,$ where the first equality follows from (\ref{3a}), while the second equality follows from the Bahadur representation of sample 
quantile. 
$\hat{\beta}_{n1}(\alpha),\ldots,\hat{\beta}_{np}(\alpha)$ are consistent estimates of the slope parameters $\beta_1\ldots,\beta_p.$
The slope components of regression quantile are asymptotically independent of the intercept component
$\hat{\beta}_{n0}(\alpha).$ 

The solution of (\ref{RQ})  
minimizes the $(\alpha, 1-\alpha)$ convex combination of residuals $(Y_i-\mathbf x_i^{*\top}\mathbf b)$ over $\mathbf b\in \mathbb R^{p+1},$ 
where the choice of $\alpha$ depends on the balance between underestimating and overestimating the respective losses $Y_i.$ The increasing 
$\alpha\nearrow 1$  reflects a greater concern about underestimating  $Y,$ comparing to overestimating.
A useful functional of the regression quantile is
 the \textit{averaged regression} $\alpha$-\textit{quantile}, the
 weighted mean of components of $\widehat{\boldgreek\beta}_n(\alpha), \; 0\leq\alpha\leq 1$: 
\begin{equation}\label{22x}
\bar{B}_n(\alpha)=\overline{\mathbf x}_n^{*\top}\widehat{\boldgreek\beta}_n(\alpha)=\widehat{\beta}_{n0}(\alpha)
+\frac 1n\sum_{i=1}^n\sum_{j=1}^p x_{ij}\widehat{\beta}_j(\alpha), \quad \overline{\mathbf x}_n^{*}=\frac 1n\sum_{i=1}^n\mathbf x_i^* 
\end{equation}
As shown in \cite{Averaged}, the $\bar{B}_n(\alpha)-\beta_0-\bar{\mathbf x}_n^{\top}\boldgreek\beta$ is asymptotically equivalent to the $[n\alpha]$-quantile 
$e_{n:[n\alpha]}$ of the model errors, if they are identically distributed. 

\subsection{Two-step regression quantile}
The two-step regression quantile
was introduced in \cite{Sankhya2005} and later studied e.g. in \cite{Aplimat}, where it is shown that it is asymptotically equivalent to the ordinary 
$\alpha$-regression quantile.
The two-step regression $\alpha$-quantile 
combines the rank-estimator (R-estimator) $\widetilde{\boldgreek\beta}_{nR}$ of the slope components 
$\boldgreek\beta$ with the $[n\alpha]$ order statistics of the  residuals $Y_i-\mathbf x_i^{\top}\widetilde{\boldgreek\beta}_{nR},  \;  i=1,\ldots,n.$

 The two-step regression quantile first estimates the slope 
components $\boldgreek\beta$ by means of an R-estimate $\widetilde{\boldgreek\beta}_{nR}(\lambda)\in{\R}^{p},$ defined
as a minimizer of the Jaeckel's measure of the rank dispersion \cite{Jaeckel1972} with a fixed $\lambda\in(0,1)$:
\begin{equation}\label{Jaeckel} 
\sum_{i=1}^n(Y_i-\mathbf x_i^{\top}\mathbf b)[a_i(\lambda,\mathbf b)-\bar{a}_{n}(\lambda)]=\min 
\end{equation}
with respect to $\mathbf b=(b_1,b_2,\ldots,b_p)^{\top}\in{\R}^{p}.$   
 The  notation in (\ref{Jaeckel}) means: 
$$a_i(\lambda,\mathbf b)=\left\{\begin{array}{lll}
0  &\ldots&\quad R_{ni}(Y_i-\mathbf x_i^{\top}\mathbf b)<n\lambda\\[1mm]
R_i-n\lambda \quad &\ldots&\quad n\lambda\leq R_{ni}(Y_i-\mathbf x_i^{\top}\mathbf b)<n\lambda+1\\[1mm]
1                 &\ldots&\quad n\lambda+1\leq R_{ni}(Y_i-\mathbf x_i^{\top}\mathbf b),\\
\end{array}\right. .
$$
 Here $R_{ni}(Y_i-\mathbf x_i^{\top}\mathbf b), i=1,\ldots,n$ are the ranks of the residuals, and $a_i(\lambda,\mathbf b)$ are known as 
\textit{H\' ajek's rank scores} (see \cite{Hajek}). Note that  
 $\bar{a}_{n}(\lambda)=\frac 1n \sum_{i=1}^n a_i(\lambda,\mathbf b)$ is constant in $\mathbf b$,  as an average of the rank scores.
The minimization (\ref{Jaeckel}) can be rewritten as
\begin{equation}\label{Jaeckel2}
\sum_{i=1}^n\left(Y_i-\bar{Y}_n-(\mathbf x_i-\bar{\mathbf x}_n)^{\top}\mathbf b\right)a_i(\lambda,\mathbf b)=\min.
\end{equation}2
It implies that the solution of (\ref{Jaeckel}) is invariant to the intercept, which is a nuisance component.
 The solution of (\ref{Jaeckel}) and (\ref{Jaeckel2}) is the R-estimator $\widetilde{\boldgreek\beta}_{nR}(\lambda)$ of  
$\boldgreek\beta=(\beta_1,\ldots,\beta_p)^{\top}$, generated by the following score function $\varphi_{\lambda}: (0,1)\mapsto \R^{1}$:
\begin{equation}\label{score} 
 \varphi_{\lambda}(u)+(1-\lambda)=\left\{\begin{array}{lll}0 \; & \ldots \; &0\leq u<\lambda\\
                                                 1    & \ldots    &\lambda\leq u\leq 1. \end{array}\right. .
\end{equation}
Generally, as the score function we can use another nondecreasing square integrable function on $(0,1)$. 

 By \cite{Sen},  $\widetilde{\boldgreek\beta}_{nR}(\lambda)$  consistently estimates $\boldgreek\beta$ under the following conditions on $F$ 
and on $\mathbf X_n$:
\begin{description}
\item[(F1)] The distribution function $F$ has a continuous density $f$ with a positive an finite Fisher information $\mathcal I(f)$.
\item[(X1)] 
Assume that, as $\ny$,
\begin{eqnarray}\label{Noether2}
 &&n^{-1}\mathbf V_n=O_p(1) \quad \mbox{ where }  \;
\mathbf V_n=\sum_{i=1}^n (\mathbf x_{ni}-\bar{\mathbf x}_n)(\mathbf x_{ni}-\bar{\mathbf x}_n)^{\top}, \\  
&& \max_{1\leq i\leq n}\|\mathbf x_{ni}-\bar{\mathbf x}_n\|=o(n^{1/4}), \quad     \bar{\mathbf x}_n=n^{-1}\sum_{i=1}^n\mathbf x_{ni}.\nonumber
\end{eqnarray}
 Moreover, we assume that $\mathbf V_n$ satisfies 
\begin{equation}\label{Noether}
 \lim_{\ny}\max_{1\leq i\leq n}(\mathbf x_{ni}-\bar{\mathbf x}_n)^{\top}\mathbf V_n^{-1}(\mathbf x_{ni}-\bar{\mathbf x}_n)=0.
\end{equation}
\end{description} 
Under  conditions (F1) and (X1),  the R-estimator $\widetilde{\boldgreek\beta}_{nR}=
\widetilde{\boldgreek\beta}_{nR}(\lambda)$
admits the following asymptotic representation, as $\ny$  (see e.g. \cite{Sen} for the proof):
\begin{equation}\label{Sen6.68}
 \widetilde{\boldgreek\beta}_{nR}-\boldgreek\beta=(f(Q(\lambda))^{-1}\mathbf V_n^{-1}
\sum_{i=1}^n(\mathbf x_{ni}-\bar{\mathbf x}_n)\Big(I[Z_{ni}>Q(\lambda)]-(1-\lambda)\Big) +o_p(n^{-1/2}),
\end{equation}
hence $\|n^{1/2}(\widetilde{\boldgreek\beta}_{nR}-\boldgreek\beta)\|=O_p(1).$

The intercept component of the two-step regression\\   $\alpha$-quantile 
is  defined as the $[n\alpha]$-quantile of the residuals $Y_i-\mathbf x_i^{\top}\widetilde{\boldgreek\beta}_{nR}(\lambda),  \; i=1,\ldots,n.$ 
Denote it as
$\widetilde{\beta}_{nR,0}(\alpha), $  hence 
$$\widetilde{\beta}_{nR,0}(\alpha)=\Big(Y_i-\mathbf x_i^{\top}\widetilde{\boldgreek\beta}_{nR}(\lambda)\Big)_{n:[n\alpha]}$$
and we define the two-step $\alpha$-regression quantile as the vector in $\R_{p+1}$
\begin{equation}\label{twoste}
\widetilde{\boldgreek\beta}_n(\alpha)=\left(\widetilde{\beta}_{nR,0}(\alpha), (\widetilde{\boldgreek\beta}_{nR}(\lambda))^{\top}\right)^{\top}.
\end{equation}
Hence,  the averaged two-step regression $\alpha$-quantile equals to 
\begin{equation}\label{averaged}
\widetilde{B}_{n\alpha}=\widetilde{\beta}_{nR,0}(\alpha)+\bar{\mathbf x}^{\top}_n\widetilde{\boldgreek\beta}_{nR}(\lambda)
=\left(Y_i-(\mathbf x_i-\bar{\mathbf x}_n)^{\top}\widetilde{\boldgreek\beta}_{nR}(\lambda)\right)_{n:[n\alpha]}
\end{equation}
The detailed account of regression quantiles and regression rank score processes can be found in \cite{Gutenbrunner}.
The averaged two-step regression $\alpha$-quantile has been introduced in \cite{Aplimat}, where it is proven that
\begin{equation}\label{averaged2}
\widetilde{B}_{n\alpha}-\beta_0-\bar{\mathbf x}_n^{\top}\boldgreek\beta=Z_{n:[n\alpha]}+o_p(n^{-1/2}) 
= Q(\alpha) +o_p(n^{-1/2}) \; \mbox{ as } \; \ny
\end{equation}
uniformly for $\alpha\in(\varepsilon, 1-\varepsilon), \; 0<\varepsilon\leq 1/2,$ and for any fixed $\lambda\in{0,1}$. 

\section{Estimation of  the functional $\mathcal S(Q)$} 
\setcounter{equation}{0}
If  there are available independent observations $Z_1, Z_2,\ldots, Z_n$ then the linear functional $\mathcal S(Q)$  can be estimated with 
the aid of  their empirical quantile function, even if distribution function of $Z$ is unknown.
The corresponding estimate  would be the corresponding functional of the empirical quantile function.

If the observations $Z_1, Z_2,\ldots, Z_n,$ are not at disposal, we can profit from approximations (\ref{averaged}) and (\ref{averaged2})
and estimate the functional  by means of the averaged two-step regression quantile $\widetilde{B}_n(\alpha)$, which explicitly contains only the available
observations $Y_1,\ldots,Y_n$. 

 The estimate of  $\mathcal S_Z{Q}$  
  is determined up to the nuisance $\beta_0+\bar{\mathbf x}_n^{\top}\boldgreek\beta.$ 
The nuisance parameters  can be also approximated by means of $Y_{n1}, \ldots, Y_{nn}$ under the conditions (F1) and (X1), namely by 
$\bar{\mathbf x}_n^{\top}\widetilde{\boldgreek\beta}_{nR}(\lambda)$ and by $\bar{Y}_{n}$.
Indeed, by (\ref{1b}) and (\ref{df}),
$$\bar{Y}_{n}=\beta_0+\bar{\mathbf x}_{n}^{\top}{\boldgreek\beta}+\bar{Z}_{n}=\beta_0+\bar{\mathbf x}_{n}^{\top}{\boldgreek\beta}+o_p(n^{-1/2})$$
as $\ny.$

Summarizing, our estimate of the functional $\mathcal S_Z(Q)$  in the linear model is described as follows: 
We assume that the distribution of  $Z$  satisfies condition (F1), 
that $\int z dF(z)=0$,  $ \; 0<\int z^2dF(z)<\infty,$  and  that the 
matrix $\mathbf X_n $  satisfies (X1).  Then, given a fixed $\lambda \in(0,1)$,
\begin{equation}\label{4c}
\widehat{\mathcal S}_{Z}(Q)
=\mathcal S(\widetilde{B}_n)+\mathcal S(\bar{Y}_n)=\mathcal S_Z(Q)+o_p(n^{-1/2}) \;  \mbox{  as  }  \;  \ny.
\end{equation}
Moreover,
\begin{equation}\label{4d}
\left(Y_i-(\mathbf x_i-\bar{\mathbf x}_n)^{\top}\widetilde{\boldgreek\beta}_{nR}(\lambda)\right)_{n:[n\alpha]}
-\bar{Y}_n=Z_{n:[n\alpha]}+o_p(n^{-1/2}) \; \mbox{  as  } \; \ny.
\end{equation}

\section{Applications}
\setcounter{equation}{0}
The typical quantile functionals are measures of risk in various contexts, as  in finances,  in health problems, environmental and technical risks. 
The popular risk measure is the \textsl{Conditional Value-at-Risk} (or \textit{expected shortfall} ) $\sf CVaR$ equal to 
\begin{equation}\label{1a}
\sf{CVaR}_{\alpha}(Z)=\E\{Z|Z>Q(\alpha)\}=(1-\alpha)^{-1}\int_{\alpha}^1 Q(u)du=(1-\alpha)^{-1}\int_{Q(\alpha)}^{\infty}zdF(z).
\end{equation}
It has obtained applications in many areas immediately after its introduction; let us mention the management of water supplies, risk management
of the social security fund, the cash flow risk measurement for non-life insurance industry, the
financial risk in the industrial areas, operational risk in the banks, and others. There is also a rich bibliography on the subject, both from theoretical and
applications points. By (\ref{4c}), we get its estimation in the form
\begin{equation}\label{4b}
\widehat{\sf{CVaR}}_{n\alpha}(Z) ={\left\lfloor n(1-\alpha)\right\rfloor}^{-1}
\sum_{i={\left\lfloor n\alpha\right\rfloor}}^n 
\sum_{\alpha\leq\delta<1}\widetilde{B}_n(\delta)-\beta_0-\bar{\mathbf x}_n^{\top}\boldgreek\beta.
\end{equation} 
The estimation of  $\sf{CVaR}_{\alpha}(Z)$ under various experimental conditions is studied in \cite{Kalina}. A review of quantile functionals used 
in finances is given in \cite{Peracchi}.
The closely related concepts are \textit{the mean excess function} and \textit{the Lorenz curve}. 
The mean excess function is the mean excess over a threshold $\gamma$, that is 
$$e(\gamma) = \E(Z-\gamma | Z \geq \gamma) = \E(Z| Z \geq \gamma) - \gamma.$$
The Lorenz curve is  used in economics to describe the distribution of income. For a nonnegative $Z$ with the mean $\mu$, it is defined as
$$L(\alpha) = \frac{1}{\mu}\int_0^{\alpha}Q(u)du,  \; \;  0<\alpha<1.$$
\cite{Gastwirth} considers the ratio of the Lorenz curve at $\alpha$ to its value at $1-\alpha$ as a measure of the fraction of income that the lowest 
$100\alpha\%$ of the population have relative to the upper $100\alpha \%$, i.e. the curve
$$J(\alpha)=\frac{L(\alpha)}{1-L(1-\alpha)}, \; 0<\alpha<1.$$
The area  between $J(\alpha)$ and 1 can be regarded as a measure of inequality;  its size for $\alpha=0.5$ is of interest.
\cite{Staudte} propose the symmetric ratio of quantiles $$R(p) \equiv\frac{Q(\alpha/2)}{Q(1 - (\alpha/2))}$$ 
as an inequality measure of income. If $F(0)<1/2$,
then $R(\alpha)$ is scale invariant and nondecreasing in $\alpha\in(0,1)$. 
\cite{Gastwirth} and \cite{Staudte} contain a rich additional bibliography on the
matter.

The nuisance covariates of the linear model can in a combination affect the final risk or other entity of the system, though they are not of a primary inerest.
They can include the weather conditions (temperature, humidity, wind speed, 
precipitation, seasonal trends), energy price fluctuations, natural events (power outages), or industrial activity.
In finances they include 
market conditions (interest rates, overall market sentiment), liquidity shocks, etc.
The nuisance effects were considered in \cite{kha} in testing the predictive models 
or in \cite{dim} for combining two forecasts. 

\section*{Conclusion}
We propose a consistent nonparametric estimates of the quantile functionals  of variable $Z$, whose observations are not available directly, 
while we observe only the responses affected by covariates with unknown intensities. The estimates are based on the averaged two-step regression 
$\alpha$-quantile of the linear model, through an R-estimator of the slope components. The functionals of interest can be the risk measures, inequality ratios
of the income, among many others.

\end{document}